\documentclass[iop,revtex4,appendixfloats]{emulateapj}
\usepackage{subfigure}
\usepackage{upgreek}
\usepackage{amsmath}
\usepackage{amssymb}
\begin{document}

\title{The Covering Factor of Warm Dust in Weak Emission-Line Active Galactic Nuclei}

\author{Xu-Dong Zhang\altaffilmark{1,2} and Yuan Liu\altaffilmark{2}}

\altaffiltext{1}{University of Chinese Academy of Sciences,
	Beijing 100049, China}

\altaffiltext{2}{Key Laboratory of Particle Astrophysics, Institute of High Energy Physics,
	Beijing 100049, China}

\email{zhangxd@ihep.ac.cn; liuyuan@ihep.ac.cn}

\begin{abstract}
Weak emission-line active galactic nuclei (WLAGNs) are radio-quiet active galactic nuclei (AGNs) that have nearly featureless optical spectra. We investigate the ultraviolet to mid-infrared spectral energy distributions of 73 WLAGNs (0.4 $<$ \emph{z} $<$ 3) and find that most of them are similar to normal AGNs. We also calculate the covering factor of warm dust of these 73 WLAGNs. No significant difference is indicated by a KS test between the covering factor of WLAGNs and normal AGNs in the common range of bolometric luminosity. The implication for several models of WLAGNs is discussed. The super-Eddington accretion is unlikely the dominant reason for the featureless spectrum of a WLAGN. The present results favor the evolution scenario, i.e., WLAGNs are in a special stage of AGNs.

\end{abstract}


\section{Introduction}
The unified model of active galactic nuclei (AGNs) includes multiple structures around the central black hole, e.g., accretion disk, broad-line region (BLR), and dusty torus. In this model, the UV/X-ray photons from the accretion disk and corona ionize the gas in the BLR, giving rise to broad emission lines in the optical spectrum.  The dust in the torus absorbs the radiation from the central engine and re-emits it in the infrared (IR) band, leading to the IR bump in the  spectral energy distribution (SED). The AGNs would exhibit different characteristics with different orientations of the line of sight, which is the most important assumption in AGN unification (Antonucci 1993).

As a subclass of AGNs, BL Lac objects are characterized by strong radio emission, polarized continua, rapid variation, and nearly featureless spectra (Urry
$\&$ Padovani 1995). According to the unified model of AGNs, they are explained by the small angle between the jet and the line of sight (Antonucci 1993). Because of the dilution by the Doppler$-$boosted relativistic jet, there will be weak or even no emission lines. In the last decade, a handful of radio-quiet AGNs with weak emission lines (WLAGNs\footnote{Because it is still unclear whether those objects are weak emission-line quasars or radio-quiet BL Lac objects, we call them WLAGNs as a whole.}) are discovered (McDowell et al. 1995; Collinge et al. 2005; Diamond$-$Stanic et al. 2009; Shemmer et al. 2009; Plotkin et al. 2010a (P10a hereafter)). Diamond$-$Stanic et al. (2009) reported the optical polarization of seven WLAGNs and found that the largest degree of polarization of these sources is 2.43\symbol{37}, which is lower than the nominal level of high polarization (degree of polarization $\geq$ 3\symbol{37}; Impey $\&$ Tapia 1990). Heidt \& Nilsson (2011) measured the optical polarization of 25 WLAGN candidates and found 23 of them are unpolarized. Gopal-Krishna et al. (2013), Chand et al. (2014), and Kumar et al. (2015) have monitored 15 WLAGNs and found all of them display intranight optical variability. They claimed that their variability duty cycles are higher than for radio-quiet quasars and radio lobe-dominated quasars. However, Liu et al. (2015) have not detected significant microvariations in eight WLAGNs. The properties of WLAGNs are different from those of BL Lac objects, but consistent with normal AGNs in most respects except for their featureless spectra. A radiatively inefficient accretion flow could cause weak emission lines (Yuan \& Narayan 2004), but the high luminosity of WLAGNs excludes this possibility. Plotkin et al. (2010b) investigated 13 radio-quiet BL Lac candidates and found that the relativistic beaming seems not to be the cause of weak lines. However, they cannot exclude the possibility of the existence of weak jets in some of these sources.

Several explanations have been proposed for the nearly featureless optical spectra of WLAGNs. Leighly et al. (2007a,b) considered that an extremely high accretion rate in WLAGNs will induce a soft ionizing continuum, which cannot produce enough high-energy photons for the BLR. Besides that, Wu et al. (2011) assumed that there is some shielding gas with a high covering factor between the accretion disk and BLR. It would absorb the high-energy ionizing photons from the accretion disk and prevent them from generating the emission lines in the BLR. This kind of ``shielding gas'' is further considered as a geometrically thick accretion disk formed by super-Eddington accretion (Luo et al. 2015). Also, a cold accretion disk, formed by a massive black hole and a relatively low accretion rate, will not produce enough ionizing photons either (Laor $\&$ Davis 2011). In these models, the weak emission lines are caused by insufficient ionizing photons. Thus, the IR luminosity of WLAGNs might be relatively lower than that of normal quasars.

Another model for the featureless \textbf{optical} spectra of WLAGNs is based on the unusual BLR, e.g., ``anemic'' BLR (Shemmer et al. 2010). Niko{\l}ajuk \& Walter (2012) estimated the covering factor of the BLR in WLAGNs from the equivalent width (EW) of C \uppercase\expandafter{\romannumeral4} emission line and the ratio of X-ray to optical luminosity. They found that the covering factors of BLRs in WLAGNs are lower than those in normal quasars by a factor of 10. Niko{\l}ajuk \& Walter (2012) argued that WLAGNs are the consequence of a low covering factor of the BLR. WLAGNs can also be in the early stage of AGN evolution. In this stage, the radiative feedback blows off the gas in the BLR and there will be no emission lines (Hryniewicz et al. 2010; Liu $\&$ Zhang 2011; Banados et al. 2014; Meusinger \& Balafkan 2014). textbf{In the case of an unusual BLR, the covering factors of tori of WLAGNs are unaffected and expected to be consistent with those of normal quasars.

The UV to mid-infrared (MIR) SEDs of WLAGNs are generally consistent with those of normal radio-quiet AGNs (Lane et al. 2011; Wu et al. 2012; Luo et al. 2015), which are dominated by two bumps corresponding to the emission from the accretion disk and torus. We will expand the sample and focus on the UV to MIR SED, comparing the temperature of the torus of  low-redshift WLAGNs with that of high-redshift WLAGNs in Lane et al. (2011). Many works suggest that there are correlations between the covering factor of the torus and AGN properties, e.g., bolometric luminosity ($L_{\rm Bol}$) and black hole mass (\emph{M}$_{\rm BH}$) (e.g., Maiolino et al. 2007; Treister et al. 2008; Mor $\&$ Trakhtenbrot 2011; Ma $\&$ Wang 2013). This work will investigate the ratio of the MIR luminosity to the bolometric luminosity of WLAGNs, which corresponds to the warm dust covering factor of the torus (CF$_{\rm WD}$). It is insightful to test whether WLAGNs follow the same correlation of normal radio-quiet AGNs. Furthermore,  the value and distribution of CF$_{\rm WD}$ of WLAGNs will be helpful in understanding their origin. Sample selection is shown in Section 2, and then the SED and investigation of covering factor are described in Section 3. We discuss the findings in Section 4.

\section{Sample Selection\label{sec2}}

Our samples are selected from Table 3 and 6 of P10a.  P10a selected 723 optical BL Lac candidates from the Sloan Digital Sky Survey (SDSS; York et al. 2000) Data Release 7 (Abazajian et al. 2009) spectroscopic database. The selection criteria are: (1) no strong emission lines in the spectra, i.e., EW is smaller than 5 ${\rm \AA}$ in the rest frame; (2) if there is a Ca\uppercase\expandafter{\romannumeral2} H/K break, then it must be smaller than 40\symbol{37}. The sources are defined as radio-quiet if the radio to optical spectral index $\alpha_{ro}<0.2$.\footnote{$\alpha_{ro}=-\log(L_o/L_r)/5.08$, where $L_r$ and $L_o$ are the radio and optical luminosities at the rest$-$frames 5 GHz and 5000 ${\rm \AA}$, respectively.} As a result, there are 86 WLAGNs selected in table 6. In table 3, P10a listed 20 high-redshift WLAGNs (\emph{z} $>$ 2.2) with EW $<$ 10 ${\rm \AA}$ for Ly$\alpha{\rm +}$N \uppercase\expandafter{\romannumeral5} in the rest frame. Six of these sources are already included in table 6. Thus there are 100 WLAGNs selected in P10a in total.

The SDSS spectra of the sample are inspected visually. Some absorbed AGNs, stars, and the sources with low-confidence redshift are further removed from our sample. We also reject a few sources due to the limitation of our calculation method. 27 sources are removed and there are 73 WLAGNs in our final sample.

\subsection{Absorbed AGNs\label{sec21}}

The optical spectra of seven sources (SDSS J155651.00+385446.4, SDSS J083413.90+511214.6, SDSS J093544.16+124031.6, SDSS J101754.91+431201.1, SDSS J114038.91+100535.3, SDSS J161016.47+303945.9, and SDSS J165806.77+611858.9) are red and dominated by the host galaxies. Thus, we remove them from our sample as being absorbed AGNs.

\subsection{Stars\label{sec22}}

 Three sources (SDSS J095438.39+234118.2, SDSS J090107.64+384658.8, and SDSS J121929.45+471522.8) show large proper motions, i.e., 32 mas yr$_{\rm -1}$, 62 mas yr$_{\rm -1}$, and 112 mas yr$_{\rm -1}$, respectively (the proper motions data are from USNO-B1; Monet et al. 2003).  Since the proper motions of more than 99\symbol{37} of SDSS quasars are lower than 30 mas/yr (Schneider et al. 2007), we remove these three sources as stars.  SDSS J121929.45+471522.8  shows a bluer spectrum than a normal BL Lac object in the optical region. Combining this with its nearly featureless spectrum, we consider it to be a DC white dwarf. Actually, SDSS J121929.45+471522.8 is listed as a DC white dwarf in the catalog of Rebassa-Mansergas et al. (2010).

\subsection{Low-confidence Redshift\label{sec23}}

It is hard to determine the redshift and further calculate the luminosity of some WLAGNs due to the lack of emission lines. We remove four sources (SDSS J141200.04+634414.9, SDSS J110938.50+373611.6, SDSS J163322.48+422738.8, and SDSS J145514.09+235337.6) because of their low-confidence redshifts.

\subsection{The lack of data\label{sec24}}

The MIR photometric data are obtained from \emph{Wide-field ~Infrared~ Survey~ Explorer} (\emph{WISE}; Wright et al. 2010) All-Sky Data Release, but eight sources (SDSS
J090107.64+384658.8, SDSS J093437.53+262232.6, SDSS J095438.39+234118.2, SDSS J103601.12+084948.4, SDSS J112809.07+022254.6, SDSS J140916.33-000011.2, SDSS J154146.37+263100.8, and SDSS J172858.15+603512.7) are not detected by \emph{WISE}. Thus, we remove them from our sample. Furthermore, five sources (SDSS J084249.02+235204.7, SDSS J114153.34+021924.4, SDSS J121221.56+534127.9, SDSS J123743.09+630144.9, and SDSS J144803.36+240704.2) are removed due to our calculation methods (see the details in Section 3).

\begin{deluxetable}{ccc}
\tablewidth{0pt}
\tablecaption{Parameters of Defining Composite SED.}
\tablehead{
\colhead{Central Wavelength } &
\colhead{Bin Range} &
\colhead{$\lambda L_{\lambda}$} \\
\colhead{($\rm \AA$)}	&
\colhead{($\rm \AA$)}  &
\colhead{(Arbitrary Units)}	}
\startdata
 1302 &  1000 $-$  1505 & 2.19 $\pm$ 0.83\\
 1768 &  1505 $-$  1975 & 2.19 $\pm$ 0.57\\
 2251 &  1975 $-$  2479 & 2.01 $\pm$ 0.46\\
 2792 &  2479 $-$  3135 & 1.89 $\pm$ 0.32\\
 3556 &  3135 $-$  4170 & 1.54 $\pm$ 0.18\\
 4839 &  4300 $-$  5498 & 1.21 $\pm$ 0.21\\
 6493 &  5498 $-$  7450 & 0.95 $\pm$ 0.23\\
 8882 &  7850 $-$ 10253 & 0.55 $\pm$ 0.27\\
11749 & 10253 $-$ 12882 & 0.44 $\pm$ 0.15\\
14363 & 12882 $-$ 16511 & 0.49 $\pm$ 0.17\\
18826 & 16511 $-$ 23800 & 0.64 $\pm$ 0.19\\
37575 & 30000 $-$ 44995 & 0.72 $\pm$ 0.28\\
51696 & 44995 $-$ 64240 & 0.76 $\pm$ 0.33\\

\enddata
\tablecomments{Errors on
$\lambda \emph{L}_{\lambda}$ are the standard deviation of the luminosity densities in each bin.}

\end{deluxetable}

\section{SED and Covering Factor\label{sec3}}

\subsection{SED of WLAGNs\label{sec31}}

We use the photometric data from \emph{WISE}, Two Micron All Sky Survey (2MASS; Skrutskie et al. 2006), UKIRT Infrared Deep Sky Survey (UKIDSS; Hewett et al. 2006), ${Galaxy~ Evolution~ Explorer}$ (\emph{GALEX}; Martin et al. 2005), and SDSS Data Release 10 (DR10; Ahn et al. 2014) to construct the SEDs of WLAGNs. If the source is detected with less than 95\symbol{37} confidence, the flux of the aperture photometry is adopted as an upper limit. These multi-band data are not observed simultaneously. However, the amplitude of variability on the scale of years will not influence our results. The mean degree of optical variability of four WLAGNs in Plotkin et al. (2010b) is 0.09 mag, and they have been observed every year since 1998. Appendix Figure \ref{fig4} presents the UV to MIR SEDs of 73 WLAGNs, which are generally consistent with the mean SED of typical quasars in Richards et al. (2006).

The SEDs of three WLAGNs are not very consistent with the mean SED of normal AGNs.  The optical spectrum and SED of SDSS J142116.23+052252.3 in the UV band are redder than normal AGNs, which is likely to be the consequence of galaxy absorption. The SEDs of SDSS J130312.89+321911.4 and SDSS J212756.67+004745.6 peak in the IR band, which is similar to  BL Lac objects.  They may still be radio-quiet BL Lac objects. However, these outliers will not influence our main conclusions.

A SED model that includes a power law with the form $\lambda\emph{L}_{\rm \lambda}$ $\propto$ $\lambda^{\rm -\alpha_{\rm \lambda}}$ and a single-temperature blackbody is adopted to fit the UV to MIR SEDs of our WLAGNs, except for the three outliers discussed above. To compare the hot dust temperature of our WLAGNs with that of the high-redshift WLAGNs in Diamond-Stanic et al. (2009) and Lane et al. (2011), the \emph{WISE} W4 photometric data are removed from the fit to be consistent with the wavelength coverage in these two references. The two \emph{GALEX} photometric data are also rejected, since the far$-$UV band is not well represented by a single power law. The best-fit models are presented in Appendix Figure \ref{fig5} and parameters are shown in table 3. Our data points for the fit to every source are limited to between 8 and 11, which leads to large uncertainties in the best$-$fit parameters. Only 22 of 70 WLAGNs in our sample have an uncertainty of temperature lower than 15$\%$. The temperature of hot dust of these 22 WLAGNs is in the range 905 K $\leq$ \emph{T} $\leq$ 1386 K and the power-law slope is in the range 0.68 $\leq$ $\alpha_{\rm \lambda}$ $\leq$ 1.32.

We further construct a composite WLAGN UV to MIR SED of all our sources, except for SDSS J142116.23+052252.3, SDSS J130312.89+321911.4, and SDSS J212756.67+004745.6. We combine all photometric data points, excluding \emph{GALEX} and \emph{WISE} W4 as well. The monochromatic luminosity of each object is normalized by the mean monochromatic luminosity at the rest-frame 4200$-$4230 ${\rm \AA}$. The data points from 70 WLAGNs are binned into five rest-frame UV bins (1000 ${\rm \AA}-$4160 ${\rm \AA}$; SDSS), two optical bins (4300 ${\rm \AA}$ $-$ 7450 ${\rm \AA}$; 2MASS/UKIDSS), four near$-$IR bins (7850 ${\rm \AA}$ $-$ 23800 ${\rm \AA}$; \emph{WISE} W1, W2), and two MIR bins (30000 ${\rm \AA}$ $-$ 64240 ${\rm \AA}$; \emph{WISE} W3). Every bin in the same waveband has the same number of photometric data points. The standard deviation of the data points in each bin is considered as the error of $\lambda \emph{L}_{\rm \lambda}$ in the composite SED. The luminosity of the composite SED is the mean luminosity of all individual SEDs contributing to the bin, and the mean wavelength of all data points in each bin represents the final wavelength of that bin in the composite SED (see Table 1). In Figure \ref{fig1} we present the composite SED and the model. The best-fit power-law index and blackbody temperature are 0.76 $\pm$ 0.08 and 960 $\pm$ 170K respectively.

\begin{figure}
\includegraphics[width=85mm]{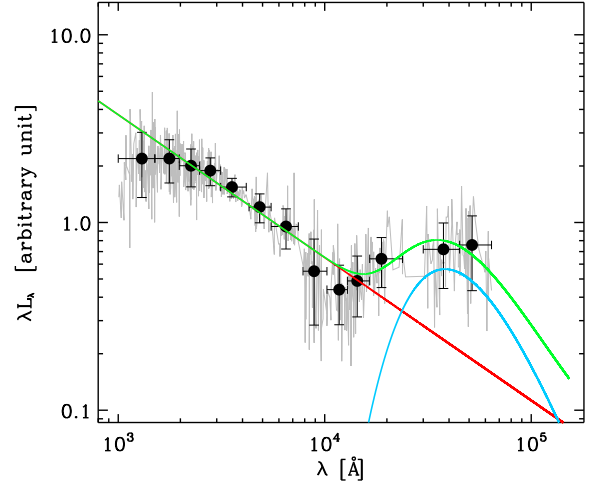}
\caption{The composite SED (filled circles) of our 70 sources. Each object is normalized at rest-frame 4215 ${\rm \AA}$. All photometric data points are combined and shown in gray. The solid red line shows the best-fit power law and the solid blue line shows the best-fit single-temperature blackbody. The sum of the two models is marked in green.\label{fig1}}
\end{figure}

\subsection{Covering Factor of warm dust in WLAGNs\label{sec32}}

We calculate the covering factor of the torus with the optical$-$ultraviolet spectra from SDSS DR10 and the IR photometric data from \emph{WISE}. All SDSS spectra are corrected for galactic extinction by using the extinction curve of Fitzpatrick (1999) and the dust map in Schlegel et al. (1998). For near-ultraviolet (NUV) continuum flux ($F_{\rm NUV}$), we adopt three relatively line-free windows 2150-2200, 3030-3100, and 4150-4250 ${\rm \AA}$ of SDSS spectra and fit them in the rest-frame with a power-law function. The NUV flux is calculated by simply integrating the best-fitting power-law curve from 2000 ${\rm \AA}$ to 4000 ${\rm \AA}$  in the rest frame. Figure~\ref{fig2} (top) shows an example of this procedure. However, there are only eight sources with low redshift (0.76$-$1.17) in our sample that could be covered by all the three windows in the rest frame. For the sources with z $>$ 2, there is just one window covered in the rest frame. To obtain reliable NUV flux, we add the \emph{J} band (if available) when fitting in the rest frame for the sources with z $>$ 1.9 (see an example in Figure~\ref{fig2} (bottom)).\footnote{We ignored \emph{GALEX} and 2MASS/UKIDSS data for the sources with z $<$ 1.9, since they are far beyond the window 2000-4000 ${\rm \AA}$ in the rest frame, which is used in the calculation of NUV flux.}  \emph{J}-band photometric data are selected from 2MASS and UKIDSS. 12 of 28 sources with \emph{z} $>$ 1.9 are detected. For the remaining 16 sources without \emph{J} band photometric data, we extend the fitting window  from 2150-2200 ${\rm \AA}$ to 2125-2225 ${\rm \AA}$ to reduce the error of parameters. We removed four sources with \emph{z} $>$ 3 from our sample, since there will be no window  covered  in the rest frame.

Our method of calculating MIR flux ($F_{\rm MIR}$) is similar to that in Ma $\&$ Wang (2013; MW13 hereafter), which uses the photometric data from \emph{WISE} W2, W3, and W4. Since the SEDs of quasars are relatively smooth in the MIR (Richards et al. 2006; Shang et al. 2011), MW13 considered it approximately as a broken power law and fixed the effective wavelength of the W3 band in the rest frame as the turning point. The MIR flux is obtained by integrating the broken power law in the rest frame  from  3 to 10 $\mu$m. The uncertainty of this approach is estimated in MW13. They applied the same method to the sample in Shang et al (2011) and compared the $F_{\rm MIR}$ with the direct integration of IR spectra over the same wavelength range. The result of their method is higher than the direct integration by 3$\%$ on average, with a scatter of 8$\%$. Thus we decreased the value of $F_{\rm MIR}$ by 3$\%$ and adopted 8$\%$ as the uncertainty of the $F_{\rm MIR}$.

\begin{figure}
\includegraphics[width=85mm]{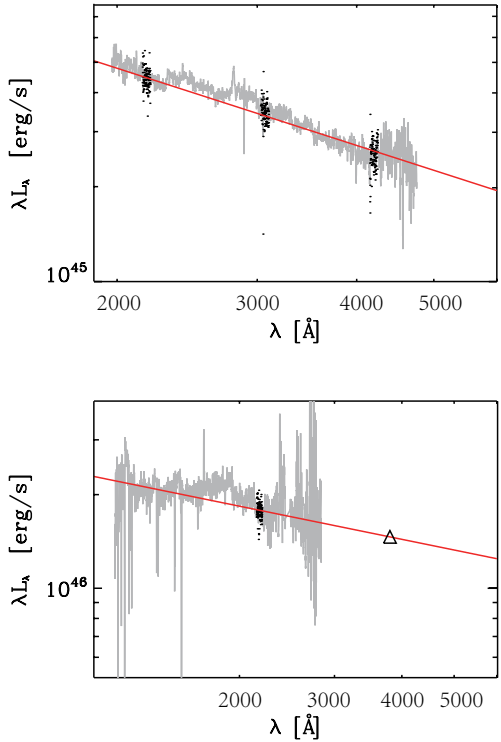}
\caption{Typical fit in the optical/NUV rest frame using a power law (red line) for the low-redshift source SDSS J131059.78+560140.1 (z = 1.28; top) and the high-redshift source SDSS J082638.59+515233.2 (z = 2.85; bottom). The SDSS spectrum (gray) and the selected windows for the fitting procedure (black) are shown. 2MASS/UKIDSS \emph{J}-band photometric data (triangles) are used for fitting high-redshift sources.\label{fig2}}
\end{figure}

The correction on NUV and MIR flux has been fully discussed in MW13. In short, $F_{\rm NUV}$  is corrected for dust extinction for the sources with NUV spectral slope $\alpha > -1.7$. The correction factor is  $3.5 \times 10^{0.322\alpha}$, which is derived from the relation between $\log(F_{\rm MIR}/F_{\rm NUV})$ and $\alpha$. Due to the contribution from the accretion disk and star formation in the MIR band, $F_{\rm MIR}$ is corrected by subtracting 0.093$F_{\rm NUV}$ from it. With the assumption of isotropic NUV and MIR emission, the MIR luminosity ($L_{\rm MIR}$) and NUV luminosity ($L_{\rm NUV}$) can be calculated from $F_{\rm MIR}$ and $F_{\rm NUV}$, respectively. We adopt bolometric correction BC $=$ 4.34 following MW13, which is calculated from the composite SED of optical blue quasars in Richard et al. (2006). Then CF$_{\rm WD}$ = $L_{\rm MIR}$/$L_{\rm Bol}$, with $L_{\rm Bol}$ = BC$\times L_{\rm NUV}$.

MW13 calculated the covering factor of the warm dust in 17,639 normal quasars with 0.76 $\leq$ z $\leq$ 1.17 and found a significant anticorrelation between CF$_{\rm WD}$ and bolometric luminosity. Our sample is too small to detect the anticorrelation. Thus we compare only the distribution of covering factors between normal quasars and the 73 WLAGNs selected above (Figure \ref{fig3}). Most of our sources are located in the luminosity range from log($L_{\rm Bol}$/erg s$^{-1}$) = 46.3$-$47.0 due to the relatively high redshift of the sample. Therefore, only the normal quasars and WLAGNs in the common luminosity range, i.e., log($L_{\rm Bol}$/erg s$^{-1}= 46.3$-$46.9$ (1667 quasars and 50 WLAGNs), are selected to compare the distribution of their covering factor by a KS test. No significant difference is detected (p-value = 0.147). We have also tried different choices of the luminosity range, e.g., log($L_{\rm Bol}$/erg s$^{-1}$) = 46.3-46.8 and 46.3-47.0. The difference is still not significant  (p-value = 0.103 and 0.107, respectively).

\begin{figure}
\includegraphics[width=85mm]{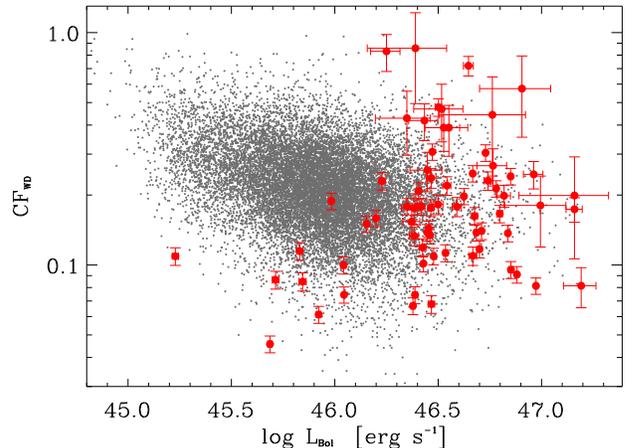}
\caption{Correlation between the covering factor of warm dust CF$_{\rm WD}$ and the bolometric luminosity $L_{\rm Bol}$. The black dots represent the 17,639 normal AGNs in MW13, while the red points represent the 73 WLAGNs in our sample.\label{fig3}}
\end{figure}

\section{Summary and Discussion\label{sec4}}

Lane et al. (2011) and Wu et al. (2012) have investigated the UV to MIR SED of 18 and 36 WLAGNs respectively, and found that the SEDs of WLAGNs  are generally consistent with the mean SED of typical quasars in the UV to MIR band. We have constructed the SEDs of 73 WLAGNs with 0.4 $\leq$ \emph{z} $\leq$ 3 and confirmed their conclusions. The SEDs of WLAGNs are significantly different from the SEDs of the blazar sequence\footnote{Blazar sequence shows a systematic trend from low-peaked, high-luminosity sources with strong emission lines to high-peaked, low-luminosity sources with weak or even no emission lines (Finke 2013). (Figure \ref{fig6})}. This suggests that the origin of their continua is likely to be the accretion disk.

The composite SED of 70 WLAGNs has been well fitted by a power law  ($\alpha_{\rm \lambda}$ $=$ 0.76) and a single-temperature black body  ($T=$ 960 K). These parameters are consistent with those of the high-redshift WLAGNs (2.7 $\leq$ $z$ $\leq$ 5.9; 0.42 $\leq$ $\alpha_{\rm \lambda}$ $\leq$ 1.21 and \emph{T} $\sim$ 1000 K) in Lane et al. (2011), suggesting that there is no significant evolution of the SEDs of WLAGNs with redshift. We have also calculated the covering factor of warm dust and found that there is  no significant difference in CF$_{\rm WD}$ between normal AGNs and WLAGNs. Plotkin et al. (2015) found that, although high-ionization emission lines (i.e., C \uppercase\expandafter{\romannumeral4}) of WLAGNs are weak, their low-ionization lines (e.g., H $\beta$, Mg \uppercase\expandafter{\romannumeral2}) remain relatively normal. Since the torus is located in the outer region of BLR, its properties (e.g., covering factors) are likely to be similar to those of normal quasars. Our results can provide some constraints on the models of WLAGNs.

\begin{figure}
\includegraphics[width=85mm]{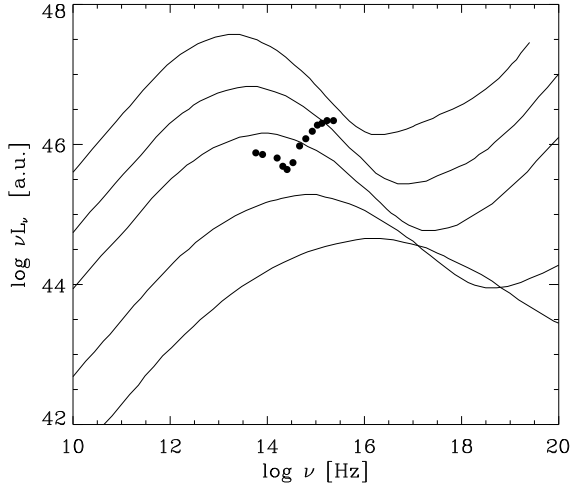}
\caption{Comparison between the composite SED of WLAGNs (filled circles) and the blazar sequence (solid curves). The SEDs of the blazar sequence are adopted from Pfrommer (2013).\label{fig6}}
\end{figure}

\subsection{Insufficient Ionization Photons? \label{sec41}}

Luo et al. (2015) assumed that the slim disk can play the role of shielding gas. However,  the illumination on the torus will be suppressed by the self-occultation of the optically and geometrically thick disk (Kawaguchi \& Mori 2011; Kawakatu \& Ohsuga 2011).\footnote{For the slim disk, $L_{\rm MIR}$/$L_{\rm Bol}$ will not represent the covering factor of warm dust, because the radiation from accretion disk is anisotropic.} Kawakatu \& Ohsuga (2011) investigated the strength of IR emission from the AGNs with sub- or super-Eddington accretion flows, by considering the anisotropic radiation flux of accretion disk. They showed that for the sub- and super-Eddington AGNs with the same covering factor as the torus, the ratio of IR luminosity to bolometric luminosity ($L_{\rm IR}/L_{\rm Bol}$) of super-Eddington AGNs is significantly less than that of sub-Eddington AGNs (see Figure 2 in Kawakatu \& Ohsuga 2011). In order to increase the observed $L_{\rm IR}/L_{\rm Bol}$ of super-Eddington AGNs to be comparable with that of sub-Eddington AGNs, the torus of super-Eddington AGN should have a large scale height. However, under the radiative feedback from the super-Eddington accretion disk, the scale height of a torus will be even smaller than that of sub-Eddington ones (combining the anisotropic radiation of super-Eddington accretion in Kawakatu \& Ohsuga [2011] and the equation of radial motion in Liu \& Zhang [2011]). Actually, Castell{\'o}-Mor et al. (2016) constructed a sample of AGNs with  similar luminosities and found that $L_{\rm IR}$/$L_{\rm Bol}$ of super-Eddington AGNs are significantly smaller than that of sub-Eddington AGNs. However, our results show the mean $L_{\rm MIR}$/$L_{\rm Bol}$ of WLAGNs and normal AGNs are 0.23 and 0.19 in the luminosity range from log($L_{\rm Bol}$/erg s$^{-1}$) $=$ 46.3 to 47.0, respectively. The strength of the IR hump in WLAGNs is even slightly higher, though the difference is not significant.

Leighly et al. (2007a,b) considered that the featureless spectrum is a consequence of a soft continuum induced by the super-Eddington accretion. Plotkin et al. (2015) and Shemmer \& Lieber (2015) calculated the Eddington ratios of 10 WLAGNs in total (0.3 $\leq$ $L$/$L_{\rm Edd}$ $\leq$ 1.3), based on the FWHM of the H$\beta$ line and the monochromatic luminosity at rest-frame 5100 ${\rm \AA}$. However, only two of them are super-Eddington and the uncertainties of Eddington ratios are large (errors of log($L$/$L_{\rm Edd}$) are larger than 0.3 dex in Shemmer\& Lieber [2015]). Niko{\l}ajuk \& Walter (2012) calculated the Eddington ratios of 76 WLAGNs from FWHM(C \uppercase\expandafter{\romannumeral4}). Only 12 sources are super-Eddington ones and the errors are large as well. Table 2 shows the $L_{\rm MIR}$/$L_{\rm Bol}$ of 5 common WLAGNs in our sample. Even for the source (SDSS J144741.76-020339.1) with the highest Eddington ratio ($L$/$L_{\rm Edd}$ $=$ 1.3), $L_{\rm MIR}$/$L_{\rm Bol}$  is still normal. It seems that the accretion rate of WLAGNs is relatively higher than than of a normal AGN, but this is \textbf{unlikely} the dominant reason for the featureless spectrum of WLAGNs.

\begin{deluxetable}{rcc}
\tablewidth{0pt}
\tablecaption{Eddington Ratios and Covering Factors.}
\tablehead{
\colhead{Name } &
\colhead{$L$/$L_{\rm Edd}$} &
\colhead{$L_{\rm MIR}$/$L_{\rm Bol}$} \\
\colhead{SDSS J}	&
\colhead{}  &
\colhead{}	}
\startdata
083650.86+142539.0 & $0.87^{+1.36}_{-0.65}$  & 0.31$\pm$ 0.02 \\
132138.86+010846.3 & $0.63^{+0.23}_{-0.37}$  & 0.07$\pm$ 0.01 \\
141141.96+140233.9 & $0.34^{+0.42}_{-0.17}$  & 0.18$\pm$ 0.01 \\
141730.92+073320.7 &  0.92$\pm$ 0.50         & 0.11$\pm$ 0.01 \\
144741.76-020339.1 & $1.30^{+0.17}_{-0.78}$  & 0.14$\pm$ 0.01 \\

\enddata
\tablecomments{The values of $L$/$L_{\rm Edd}$ here are obtained from Plotkin et al. (2015).}
\end{deluxetable}

A cold accretion disk, proposed by Laor $\&$ Davis (2011), will induce a SED with a steep fall in UV. This type of SED is observed in SDSS J094533.99+100950.1 by Hryniewicz et al. (2010). We have not found this special shape of SED in our sample. In fact, Plotkin et al. (2015) estimated the black hole mass and Eddington ratio of SDSS J094533.99+100950.1 and some other WLAGNs from the H$\beta$ line and found that these sources do not satisfy the criteria for a cold accretion disk.

\subsection{An Unusual BLR \label{sec42}}

An unusual BLR of WLAGNs can still explain the present results. Since the scale of the torus is much larger than the accretion disk, it should form first and is likely to feed the accretion disk. Therefore, the BLR should be formed after the torus and accretion disk. Hryniewicz et al. (2010) proposed a phase ($\sim$ $10^{3}$ yr) that exists in the early stage of AGN evolution. In this stage, the BLR has not yet fully formed. The gas close to the accretion disk that produces low-ionization emission lines is formed earlier than the gas away from the disk that produces high-ionization emission lines. This will result in a low covering factor of BLR, which is consistent with the conclusion of Niko{\l}ajuk \& Walter (2012). In the evolution scenario, the high-ionization emission lines are expected to be weak while the low-ionization lines should be normal, which is confirmed by Plotkin et al. (2015) as well.  Hryniewicz et al. (2010) also estimated the duty cycle of WLAGNs and found that it is consistent with the fraction of WLAGNs in SDSS quasars, though the sample is not complete. If WLAGNs are indeed in a special stage of AGNs, their fraction should be the function of redshift. A future large sample of WLAGNs with well known incompleteness could examine this evolution scenario more quantitatively.

A larger sample of WLAGNs will be helpful in clarifying the correlations of CF$-$\emph{L}$_{\rm Bol}$ and CF$-$M$_{\rm BH}$, if the  M$_{\rm BH}$ could be determined, e.g., from the H$\beta$ line.

\acknowledgements We thank X.-C. Ma for helpful discussions and the referee for constructive suggestions. This work is supported by the National
Natural Science Foundation of China under grant Nos. 11103019, 11573027, 11133002, and 11103022.

\clearpage
\LongTables
\begin{deluxetable}{rcclcccc}
\tabletypesize{\scriptsize}
\tablewidth{0pt}
\tablecaption{Primary Data of our Sample.}
\tablehead{
\colhead{Name}           & \colhead{Redshift}      &
\colhead{$\alpha_{\rm \lambda}$\tablenotemark{a}}  &     \colhead{$\rm T$\tablenotemark{a}} &
\colhead{CF$_{\rm WD}$}          & \colhead{MIR Flux}  &
\colhead{NUV Flux}          & \colhead{Bolometric Luminosity} \\
\colhead{SDSS J} &        \colhead{$z$} &
\colhead{} &        \colhead{K} &
\colhead{}       &        \colhead{$F_{\rm MIR}$\tablenotemark{b}} &
\colhead{$F_{\rm NUV}$\tablenotemark{b}} &
\colhead{log($L_{\rm Bol}$/$(\rm erg\,s^{-1}))$}    }
\startdata
001741.84$-$105613.2&1.806 &  1.19   & 1200    & 0.18$\pm$ 0.02& 2.56$\pm$0.23& 3.23$\pm$0.18& $46.50^{+ 0.02}_{-0.02}$ \\
005713.01$+$004205.9&1.546 &  0.82   & 900     & 0.13$\pm$ 0.01& 2.10$\pm$0.20& 3.63$\pm$0.19& $46.38^{+ 0.02}_{-0.02}$ \\
025716.22$-$001433.5&1.912 &  1.00   & 1200    & 0.18$\pm$ 0.03& 1.64$\pm$0.15& 2.15$\pm$0.28& $46.38^{+ 0.05}_{-0.06}$ \\
025743.73$+$011144.5&1.712 &  1.02   & 1300    & 0.18$\pm$ 0.01& 2.38$\pm$0.21& 3.07$\pm$0.04& $46.42^{+ 0.01}_{-0.01}$ \\
075331.84$+$270415.3&1.658 &  1.20   & 1100    & 0.14$\pm$ 0.01& 2.12$\pm$0.20& 3.54$\pm$0.05& $46.45^{+ 0.01}_{-0.01}$ \\
080906.88$+$172955.2&2.951 &  1.30   & 1200    & 0.57$\pm$ 0.22& 6.21$\pm$0.52& 2.49$\pm$0.94& $46.91^{+ 0.14}_{-0.20}$ \\
081250.80$+$522530.8&1.149 &  0.93   & 900     & 0.11$\pm$ 0.01& 4.45$\pm$0.43& 9.40$\pm$0.15& $46.48^{+ 0.01}_{-0.01}$ \\
082638.59$+$515233.2&2.850 &  1.17   & 1400    & 0.34$\pm$ 0.04&16.03$\pm$1.37&11.02$\pm$0.97& $47.51^{+ 0.04}_{-0.04}$ \\
082722.73$+$032755.9&2.023 &  1.16   & 1100    & 0.30$\pm$ 0.02& 5.48$\pm$0.47& 4.16$\pm$0.07& $46.73^{+ 0.01}_{-0.01}$ \\
083304.74$+$415331.3&2.329 &  1.03   & 700     & 0.43$\pm$ 0.13& 2.28$\pm$0.19& 1.23$\pm$0.36& $46.35^{+ 0.11}_{-0.15}$ \\
083330.56$+$233909.2&2.417 &  0.69   & 1100    & 0.72$\pm$ 0.07& 6.96$\pm$0.57& 2.23$\pm$0.12& $46.65^{+ 0.02}_{-0.02}$ \\
083650.86$+$142539.1&1.745 &  0.89   & 1200    & 0.31$\pm$ 0.02& 4.40$\pm$0.38& 3.30$\pm$0.05& $46.47^{+ 0.01}_{-0.01}$ \\
084424.24$+$124546.5&2.505 &  1.12   & 1400    & 0.17$\pm$ 0.02& 5.01$\pm$0.45& 6.65$\pm$0.60& $47.16^{+ 0.04}_{-0.04}$ \\
085025.60$+$342750.9&1.39  &  0.72   & 1100    & 0.13$\pm$ 0.01& 3.29$\pm$0.31& 5.64$\pm$0.15& $46.46^{+ 0.01}_{-0.01}$ \\
085701.00$+$234239.5&1.522 &  0.95   & 900     & 0.18$\pm$ 0.01& 2.70$\pm$0.24& 3.49$\pm$0.04& $46.35^{+ 0.00}_{-0.00}$ \\
090703.92$+$410748.3&2.672 &  1.37   & 1200    & 0.86$\pm$ 0.36& 3.59$\pm$0.30& 0.97$\pm$0.40& $46.39^{+ 0.15}_{-0.23}$ \\
090843.25$+$285229.8&0.932 &  1.00   & 1200    & 0.07$\pm$ 0.01& 1.88$\pm$0.19& 5.82$\pm$0.03& $46.04^{+ 0.00}_{-0.00}$ \\
092312.75$+$174452.8&2.260 &  1.06   & 900     & 0.18$\pm$ 0.06& 4.57$\pm$0.45& 5.84$\pm$1.92& $46.99^{+ 0.12}_{-0.17}$ \\
092839.08$+$533152.2&0.991 &  0.46   & 600     & 0.10$\pm$ 0.01& 2.15$\pm$0.21& 4.97$\pm$0.13& $46.04^{+ 0.01}_{-0.01}$ \\
095041.07$+$233159.6&2.114 &  1.25   & 1200    & 0.83$\pm$ 0.15& 4.46$\pm$0.37& 1.24$\pm$0.20& $46.25^{+ 0.07}_{-0.08}$ \\
095125.90$+$504559.7&1.356 &  0.64   & 400     & 0.07$\pm$ 0.01& 1.79$\pm$0.19& 6.09$\pm$0.22& $46.47^{+ 0.02}_{-0.02}$ \\
095802.93$+$062634.5&0.967 &  0.71   & 1500    & 0.06$\pm$ 0.01& 1.06$\pm$0.12& 4.00$\pm$0.11& $45.92^{+ 0.01}_{-0.01}$ \\
100517.54$+$331202.8&1.786 &  0.83   & 900     & 0.18$\pm$ 0.02& 3.17$\pm$0.28& 4.10$\pm$0.21& $46.59^{+ 0.02}_{-0.02}$ \\
100902.83$+$621354.8&1.373 &  0.79   & 900     & 0.23$\pm$ 0.02& 3.40$\pm$0.30& 3.40$\pm$0.12& $46.23^{+ 0.01}_{-0.02}$ \\
101353.46$+$492758.1&1.636 &  0.69   & 500     & 0.12$\pm$ 0.01& 3.32$\pm$0.31& 6.54$\pm$0.16& $46.70^{+ 0.01}_{-0.01}$ \\
101849.78$+$271914.9&2.603 &  1.00   & 700     & 0.47$\pm$ 0.13& 2.81$\pm$0.24& 1.38$\pm$0.37& $46.52^{+ 0.10}_{-0.14}$ \\
102609.92$+$253651.2&2.318 &  1.01   & 700     & 0.39$\pm$ 0.08& 3.17$\pm$0.27& 1.87$\pm$0.36& $46.53^{+ 0.08}_{-0.09}$ \\
103736.09$+$342730.8&1.494 &  0.77   & 900     & 0.15$\pm$ 0.01& 1.52$\pm$0.14& 2.33$\pm$0.04& $46.15^{+ 0.01}_{-0.01}$ \\
103806.71$+$382626.5&1.372 &  0.95   & 900     & 0.16$\pm$ 0.01& 2.20$\pm$0.20& 3.19$\pm$0.03& $46.20^{+ 0.00}_{-0.00}$ \\
111401.31$+$222211.5&2.121 &  0.56   & 700     & 0.08$\pm$ 0.02& 3.79$\pm$0.42&10.74$\pm$1.91& $47.19^{+ 0.07}_{-0.09}$ \\
111642.81$+$420324.9&2.526 &  0.97   & 600     & 0.27$\pm$ 0.05& 3.05$\pm$0.27& 2.63$\pm$0.43& $46.77^{+ 0.07}_{-0.08}$ \\
113413.47$+$001041.9&1.484 &  0.85   & 900     & 0.14$\pm$ 0.01& 4.81$\pm$0.45& 8.04$\pm$0.16& $46.68^{+ 0.01}_{-0.01}$ \\
113747.64$+$391941.5&2.433 &  0.94   & 600     & 0.39$\pm$ 0.10& 2.98$\pm$0.25& 1.76$\pm$0.41& $46.55^{+ 0.09}_{-0.12}$ \\
113900.55$-$020140.0&1.908 &  0.73   & 1100    & 0.20$\pm$ 0.02& 5.08$\pm$0.45& 5.89$\pm$0.44& $46.82^{+ 0.03}_{-0.03}$ \\
114137.13$-$002729.8&1.24  &  0.52   & 1000    & 0.18$\pm$ 0.01& 5.04$\pm$0.45& 6.51$\pm$0.16& $46.40^{+ 0.01}_{-0.01}$ \\
115353.35$+$502633.9&0.868 &  0.64   & 700     & 0.08$\pm$ 0.01& 1.61$\pm$0.16& 4.37$\pm$0.18& $45.84^{+ 0.02}_{-0.02}$ \\
115637.02$+$184856.5&1.985 &  1.26   & 1100    & 0.14$\pm$ 0.01& 3.32$\pm$0.31& 5.59$\pm$0.09& $46.84^{+ 0.01}_{-0.01}$ \\
115959.71$+$410152.9&2.788 &  1.25   & 1300    & 0.21$\pm$ 0.02& 7.80$\pm$0.69& 8.58$\pm$0.07& $47.38^{+ 0.00}_{-0.00}$ \\
123618.90$+$032456.0&1.008 &  1.03   & 900     & 0.11$\pm$ 0.01& 1.46$\pm$0.14& 2.93$\pm$0.10& $45.83^{+ 0.02}_{-0.02}$ \\
124203.13$+$085048.1&1.609 &  0.63   & 500     & 0.24$\pm$ 0.02& 4.09$\pm$0.36& 3.98$\pm$0.22& $46.47^{+ 0.02}_{-0.02}$ \\
124406.73$+$494936.0&2.981 &  0.98   & 1300    & 0.44$\pm$ 0.20& 3.36$\pm$0.29& 1.75$\pm$0.78& $46.76^{+ 0.16}_{-0.26}$ \\
124745.39$+$325147.0&2.249 &  0.89   & 800     & 0.17$\pm$ 0.01& 2.71$\pm$0.24& 3.76$\pm$0.11& $46.80^{+ 0.01}_{-0.01}$ \\
125219.48$+$264053.9&1.287 &  0.68   & 1000    & 0.25$\pm$ 0.02&11.79$\pm$1.02&10.97$\pm$0.16& $46.67^{+ 0.01}_{-0.01}$ \\
130312.89$+$321911.4&0.64  & \nodata & \nodata & 0.05$\pm$ 0.00& 1.27$\pm$0.15& 6.42$\pm$0.14& $45.69^{+ 0.01}_{-0.01}$ \\
131059.78$+$560140.1&1.28  &  0.75   & 1000    & 0.21$\pm$ 0.02& 5.52$\pm$0.49& 6.10$\pm$0.16& $46.41^{+ 0.01}_{-0.01}$ \\
132130.21$+$481719.2&1.407 &  0.44   & 800     & 0.11$\pm$ 0.01& 3.18$\pm$0.30& 6.50$\pm$0.16& $46.54^{+ 0.01}_{-0.01}$ \\
132138.86$+$010846.2&1.421 &  1.02   & 1000    & 0.07$\pm$ 0.01& 1.46$\pm$0.15& 4.52$\pm$0.12& $46.39^{+ 0.01}_{-0.01}$ \\
132703.26$+$341321.7&2.558 &  1.02   & 700     & 0.23$\pm$ 0.02& 2.41$\pm$0.21& 2.41$\pm$0.10& $46.74^{+ 0.02}_{-0.02}$ \\
132809.59$+$545452.8&2.098 &  1.11   & 1100    & 0.25$\pm$ 0.03& 6.93$\pm$0.61& 6.50$\pm$0.70& $46.96^{+ 0.04}_{-0.05}$ \\
133222.62$+$034739.9&1.442 &  0.61   & 300     & 0.02$\pm$ 0.00& 1.03$\pm$0.20&15.59$\pm$0.33& $46.94^{+ 0.01}_{-0.01}$ \\
134601.29$+$585820.1&1.667 &  0.92   & 900     & 0.08$\pm$ 0.01& 4.14$\pm$0.42&11.72$\pm$0.16& $46.97^{+ 0.01}_{-0.01}$ \\
140710.26$+$241853.6&1.664 &  0.83   & 1000    & 0.14$\pm$ 0.01& 3.91$\pm$0.36& 6.43$\pm$0.14& $46.71^{+ 0.01}_{-0.01}$ \\
140918.18$+$014217.5&0.937 &  0.74   & 900     & 0.19$\pm$ 0.02& 4.08$\pm$0.36& 4.99$\pm$0.09& $45.98^{+ 0.01}_{-0.01}$ \\
141141.96$+$140233.9&1.741 &  1.32   & 1300    & 0.18$\pm$ 0.01& 2.50$\pm$0.22& 3.27$\pm$0.07& $46.46^{+ 0.01}_{-0.01}$ \\
141657.93$+$123431.6&2.609 &  1.44   & 1100    & 0.24$\pm$ 0.02& 3.09$\pm$0.27& 2.96$\pm$0.06& $46.85^{+ 0.01}_{-0.01}$ \\
141730.92$+$073320.7&1.704 &  0.99   & 1200    & 0.11$\pm$ 0.01& 2.61$\pm$0.25& 5.50$\pm$0.24& $46.67^{+ 0.02}_{-0.02}$ \\
142116.23$+$052252.3&0.682 & \nodata & \nodata & 0.09$\pm$ 0.01& 2.19$\pm$0.22& 5.85$\pm$0.21& $45.71^{+ 0.02}_{-0.02}$ \\
142601.31$+$243245.7&2.121 &  1.08   & 1200    & 0.26$\pm$ 0.05& 2.15$\pm$0.19& 1.94$\pm$0.34& $46.45^{+ 0.07}_{-0.08}$ \\
142943.64$+$385932.2&0.93  &  0.92   & 900     & 0.10$\pm$ 0.01& 6.24$\pm$0.60&14.16$\pm$0.04& $46.43^{+ 0.00}_{-0.00}$ \\
143653.39$+$053408.9&1.518 &  0.77   & 500     & 0.12$\pm$ 0.01& 2.16$\pm$0.20& 4.18$\pm$0.16& $46.42^{+ 0.02}_{-0.02}$ \\
144741.76$-$020339.1&1.427 &  1.23   & 1100    & 0.14$\pm$ 0.01& 3.23$\pm$0.30& 5.21$\pm$0.04& $46.45^{+ 0.00}_{-0.00}$ \\
145604.19$+$544253.1&1.827 &  0.78   & 1000    & 0.15$\pm$ 0.02& 1.57$\pm$0.14& 2.35$\pm$0.22& $46.37^{+ 0.04}_{-0.04}$ \\
153044.08$+$231013.5&1.405 &  0.99   & 1000    & 0.21$\pm$ 0.02&10.61$\pm$0.93&11.47$\pm$0.17& $46.78^{+ 0.01}_{-0.01}$ \\
160114.81$+$042052.7&2.125 &  1.12   & 1200    & 0.42$\pm$ 0.08& 3.39$\pm$0.29& 1.87$\pm$0.30& $46.43^{+ 0.07}_{-0.08}$ \\
160410.22$+$432614.7&1.568 &  0.82   & 800     & 0.10$\pm$ 0.01& 4.27$\pm$0.42&10.31$\pm$0.25& $46.85^{+ 0.01}_{-0.01}$ \\
161245.67$+$511816.8&1.518 &  1.05   & 1100    & 0.09$\pm$ 0.01& 4.72$\pm$0.47&11.94$\pm$0.20& $46.88^{+ 0.01}_{-0.01}$ \\
162933.60$+$253200.6&1.339 &  0.43   & 400     & 0.07$\pm$ 0.01& 1.48$\pm$0.16& 5.11$\pm$0.11& $46.38^{+ 0.01}_{-0.01}$ \\
164302.05$+$441422.4&1.65  &  0.53   & 500     & 0.16$\pm$ 0.01& 4.28$\pm$0.39& 6.07$\pm$0.20& $46.68^{+ 0.01}_{-0.01}$ \\
164903.28$+$304936.5&2.05  &  0.99   & 700     & 0.22$\pm$ 0.02& 2.49$\pm$0.22& 2.62$\pm$0.12& $46.54^{+ 0.02}_{-0.02}$ \\
212416.05$-$074129.9&1.402 &  1.16   & 1000    & 0.20$\pm$ 0.02& 6.90$\pm$0.61& 8.06$\pm$0.04& $46.62^{+ 0.00}_{-0.00}$ \\
212756.67$+$004745.6&0.434 & \nodata & \nodata & 0.11$\pm$ 0.01& 2.70$\pm$0.26& 5.70$\pm$0.20& $45.23^{+ 0.01}_{-0.02}$ \\
213742.25$-$003912.7&2.257 &  1.16   & 1100    & 0.48$\pm$ 0.04& 3.90$\pm$0.33& 1.88$\pm$0.07& $46.50^{+ 0.02}_{-0.02}$ \\
233939.48$-$103539.3&2.803 &  1.37   & 900     & 0.20$\pm$ 0.09& 4.37$\pm$0.44& 5.06$\pm$2.31& $47.16^{+ 0.16}_{-0.27}$ \\

\enddata
\tablenotetext{a}{Best-fit parameters for fitting photometric data points,using a power-law plus a blackbody component.}
\tablenotetext{b}{The unit of the flux is 10$^{-13}$ $\rm erg\,s^{-1}\,cm^{-2}$.}
\end{deluxetable}

\clearpage

\appendix
\section{The UV to MIR SEDs of WLAGNs }
\begin{figure*}[!htp]
\centering
\epsscale{.80}
\plotone{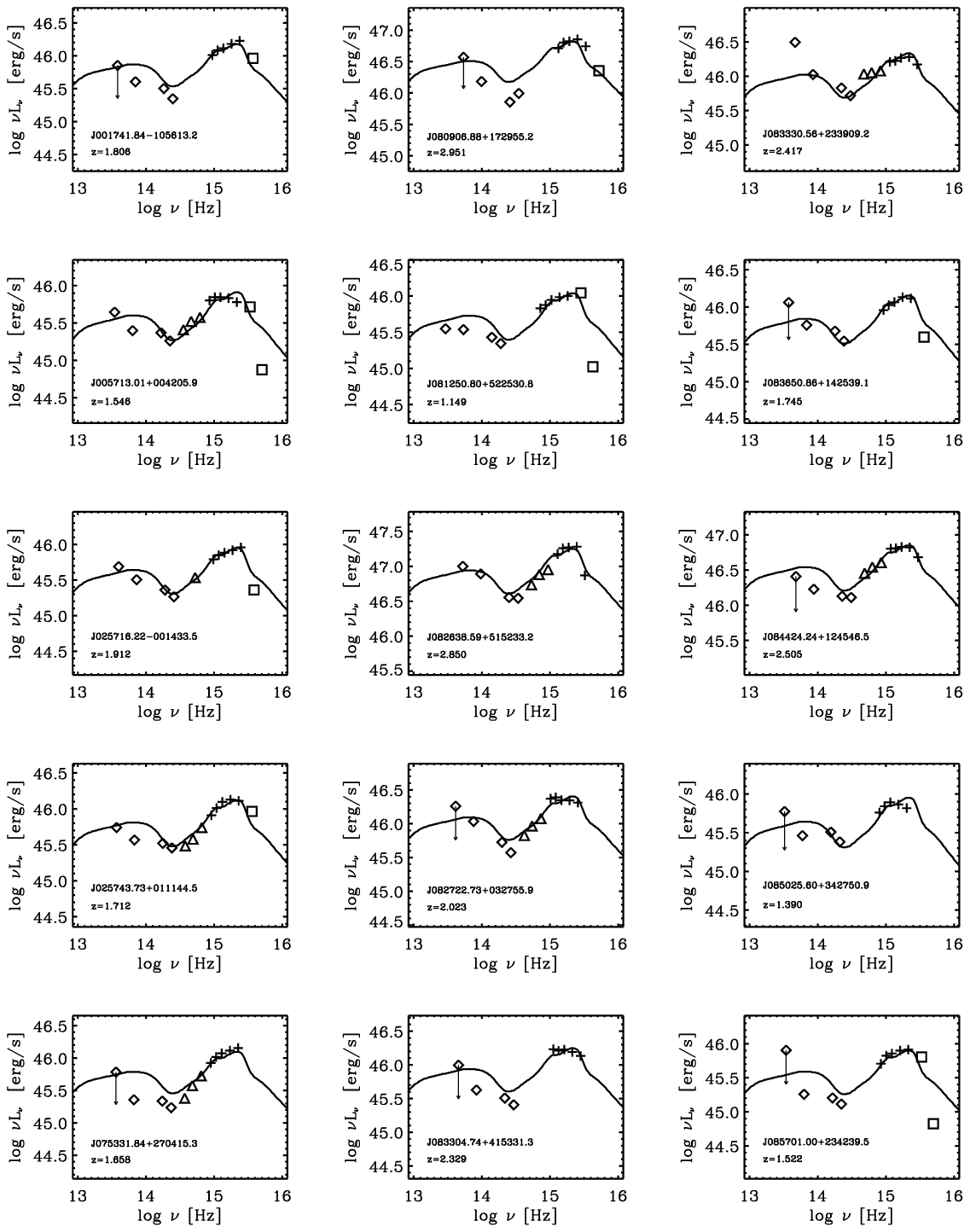}
\caption{The UV to MIR SED of 73 WLAGNs. The photometric data points are from ${GALEX}$ (squares), SDSS (plus), 2MASS or UKIDSS (triangles), and \emph{WISE} (diamonds). The solid curve is the mean SED of a normal AGN in Richards et al. (2006).\label{fig4}}
\end{figure*}

\begin{figure*}[!htp]
\centering
\epsscale{.80}
\plotone{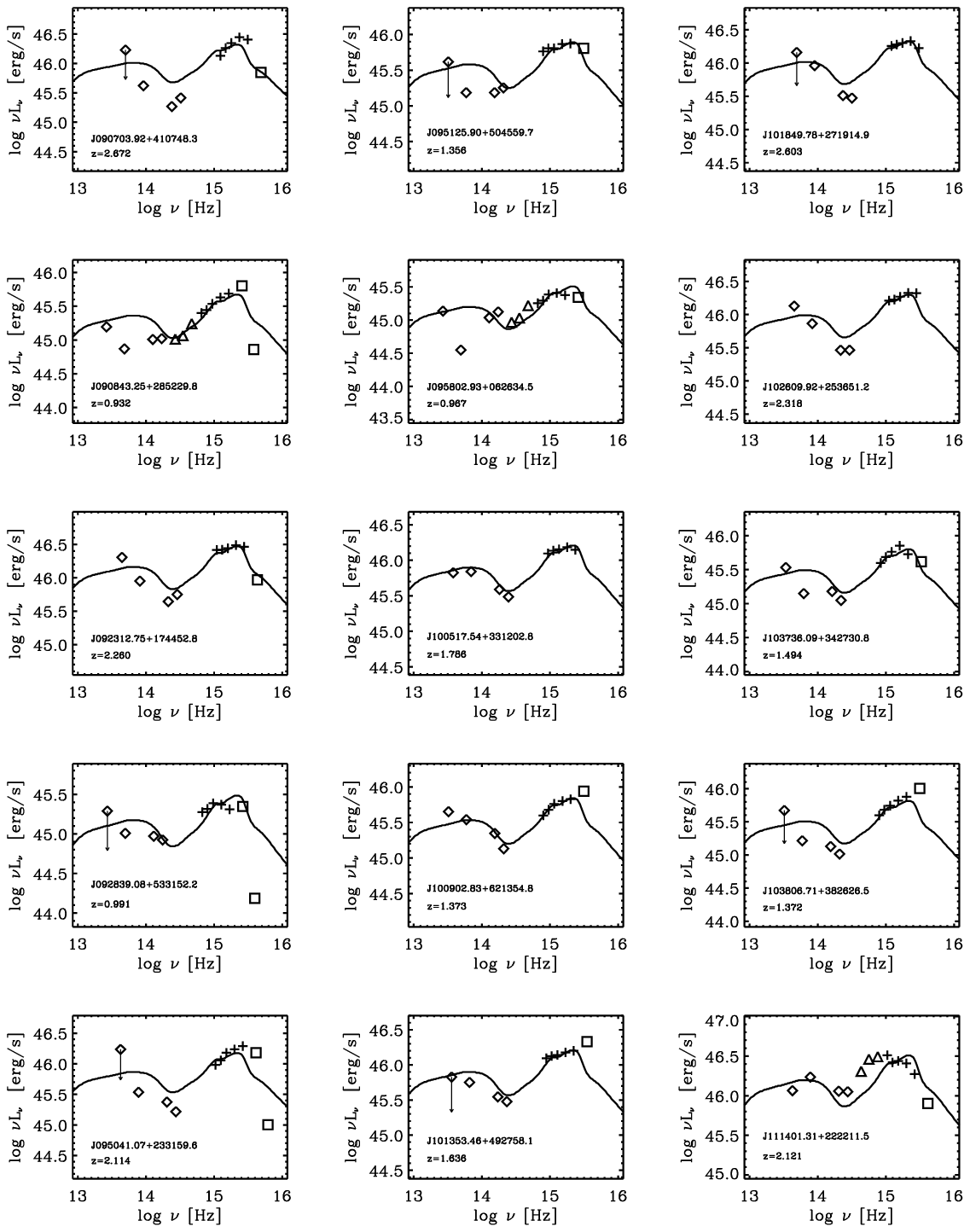}

{{\bf Figure~\ref{fig4}.} \textit {continued}}
\end{figure*}

\begin{figure*}[!htp]
\centering
\epsscale{.80}
\plotone{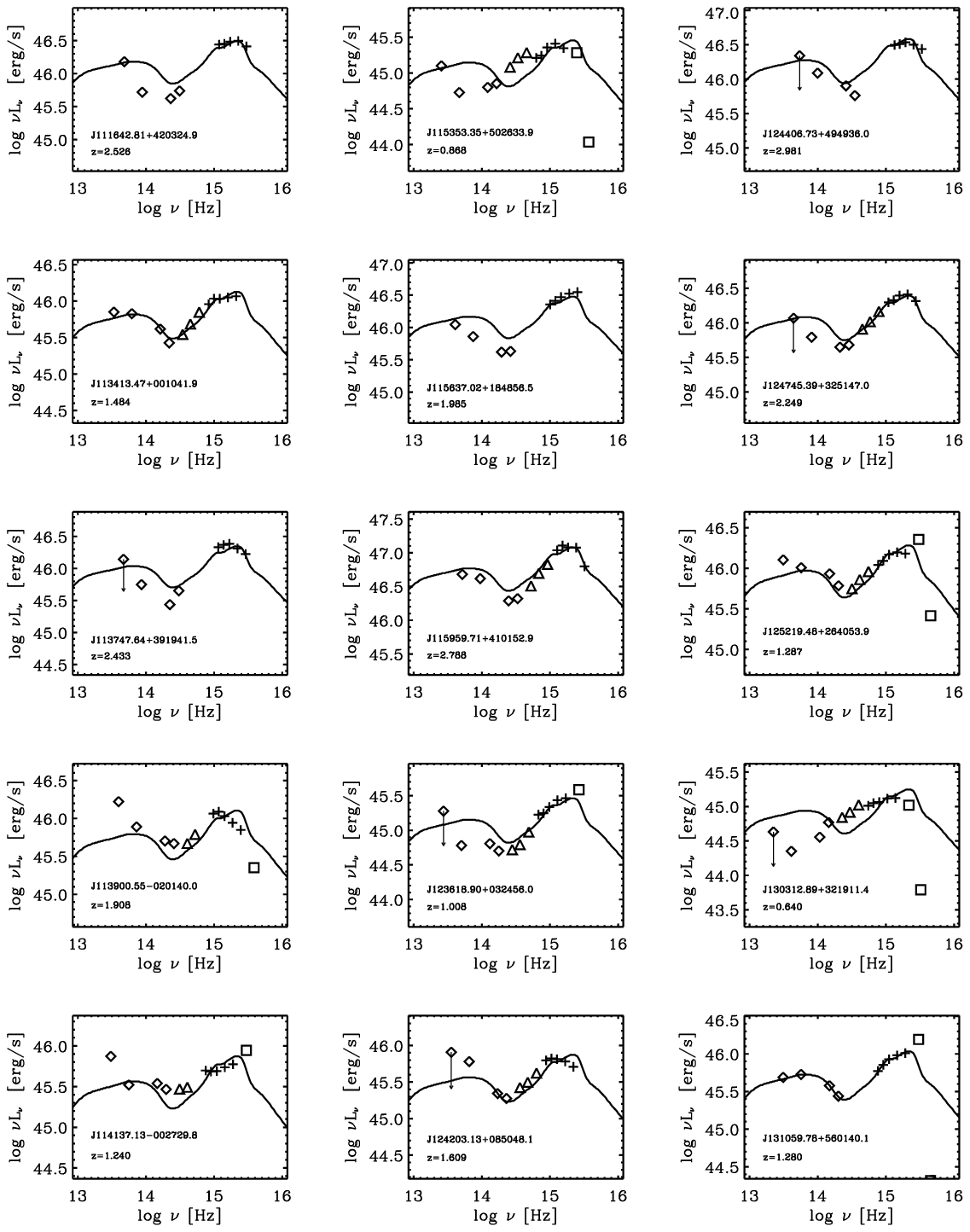}

{{\bf Figure~\ref{fig4}.} \textit {continued}}
\end{figure*}

\begin{figure*}[!htp]
\centering
\epsscale{.80}
\plotone{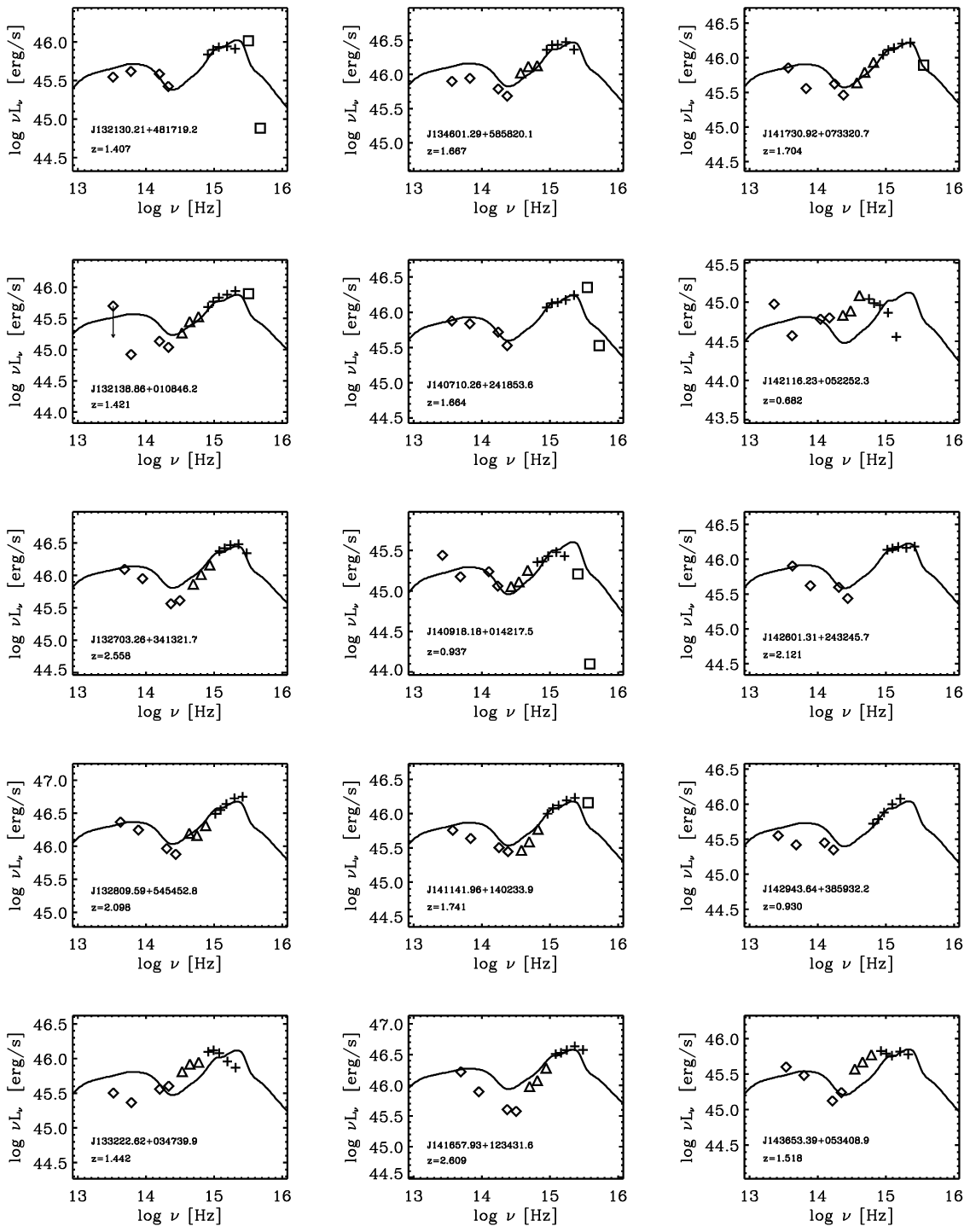}

{{\bf Figure~\ref{fig4}.} \textit {continued}}
\end{figure*}

\begin{figure*}[!htp]
\centering
\epsscale{.80}
\plotone{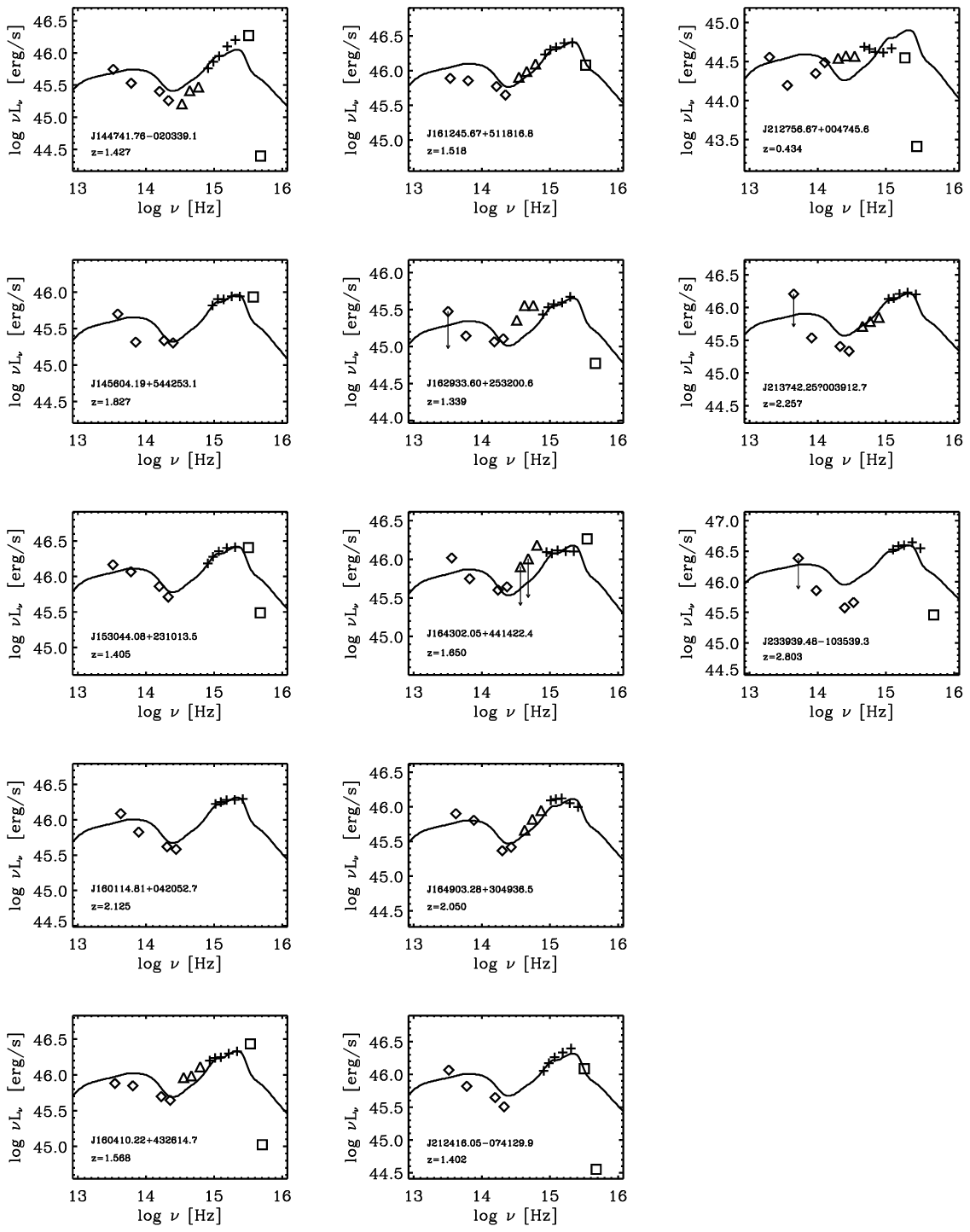}

{{\bf Figure~\ref{fig4}.} \textit {continued}}
\end{figure*}

\clearpage

\begin{figure*}[!htp]
\centering
\epsscale{.80}
\plotone{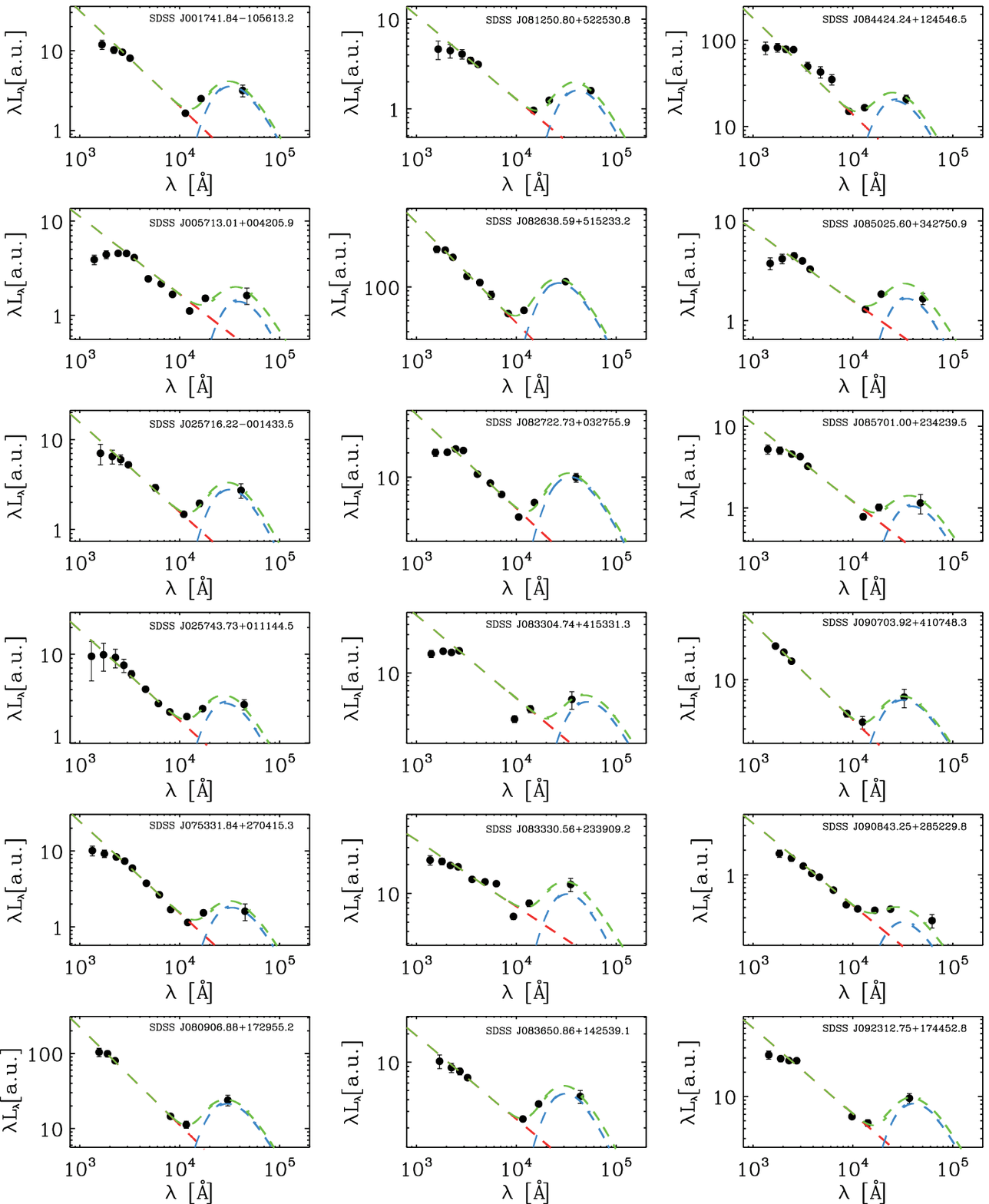}
\caption{The UV to MIR SEDs of 70 WLAGNs, fitted by a power-law (dashed-red line) plus a single-temperature blackbody (dashed-blue line).
The sum of the two components is marked as dashed-green line. The filled circles are the photometric data of SDSS, 2MASS/UKIDSS, and \emph{WISE}.\label{fig5}}
\end{figure*}

\begin{figure*}[!htp]
\centering
\epsscale{.80}
\plotone{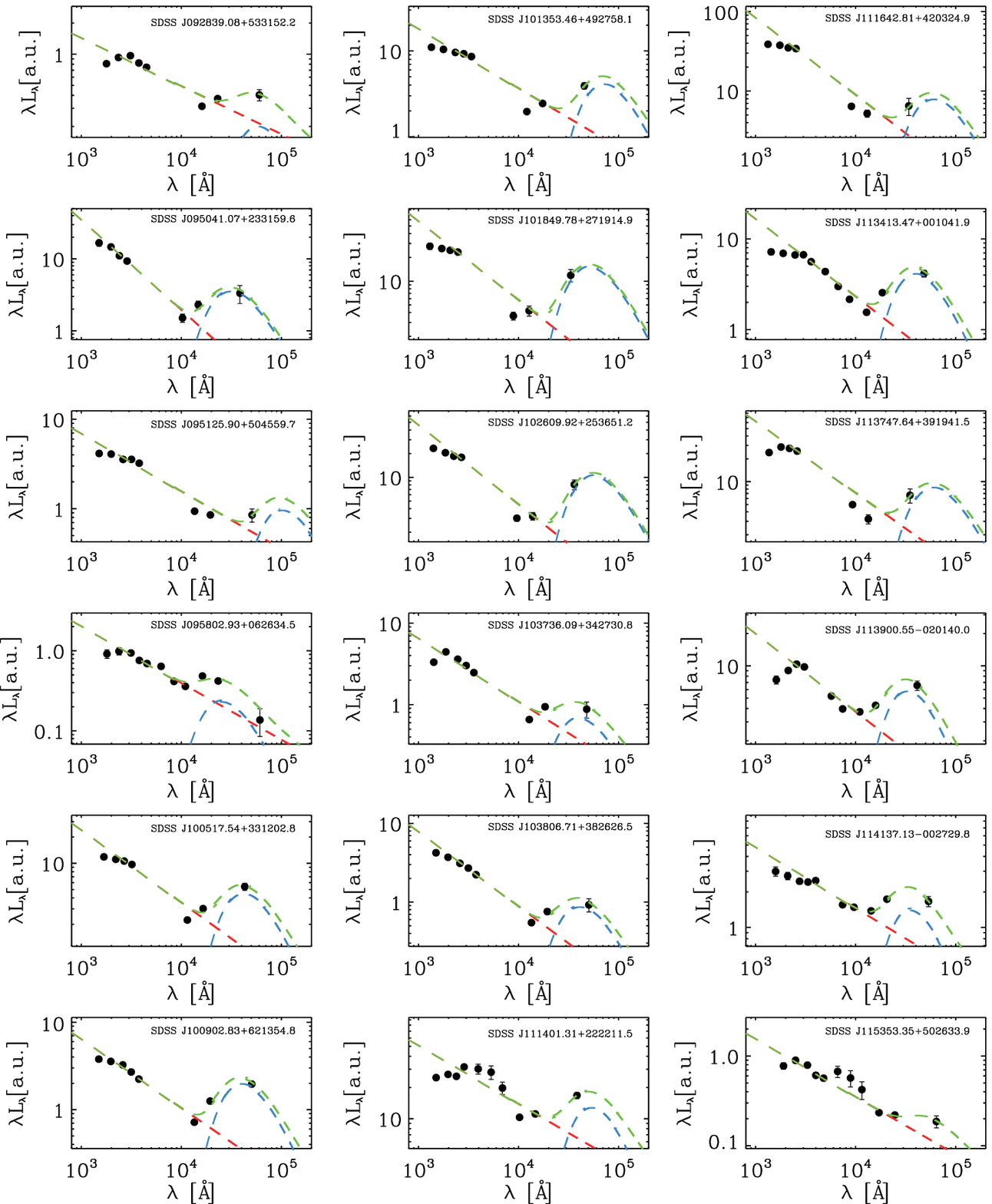}

{{\bf Figure~\ref{fig5}.} \textit {continued}}
\end{figure*}

\begin{figure*}[!htp]
\centering
\epsscale{.80}
\plotone{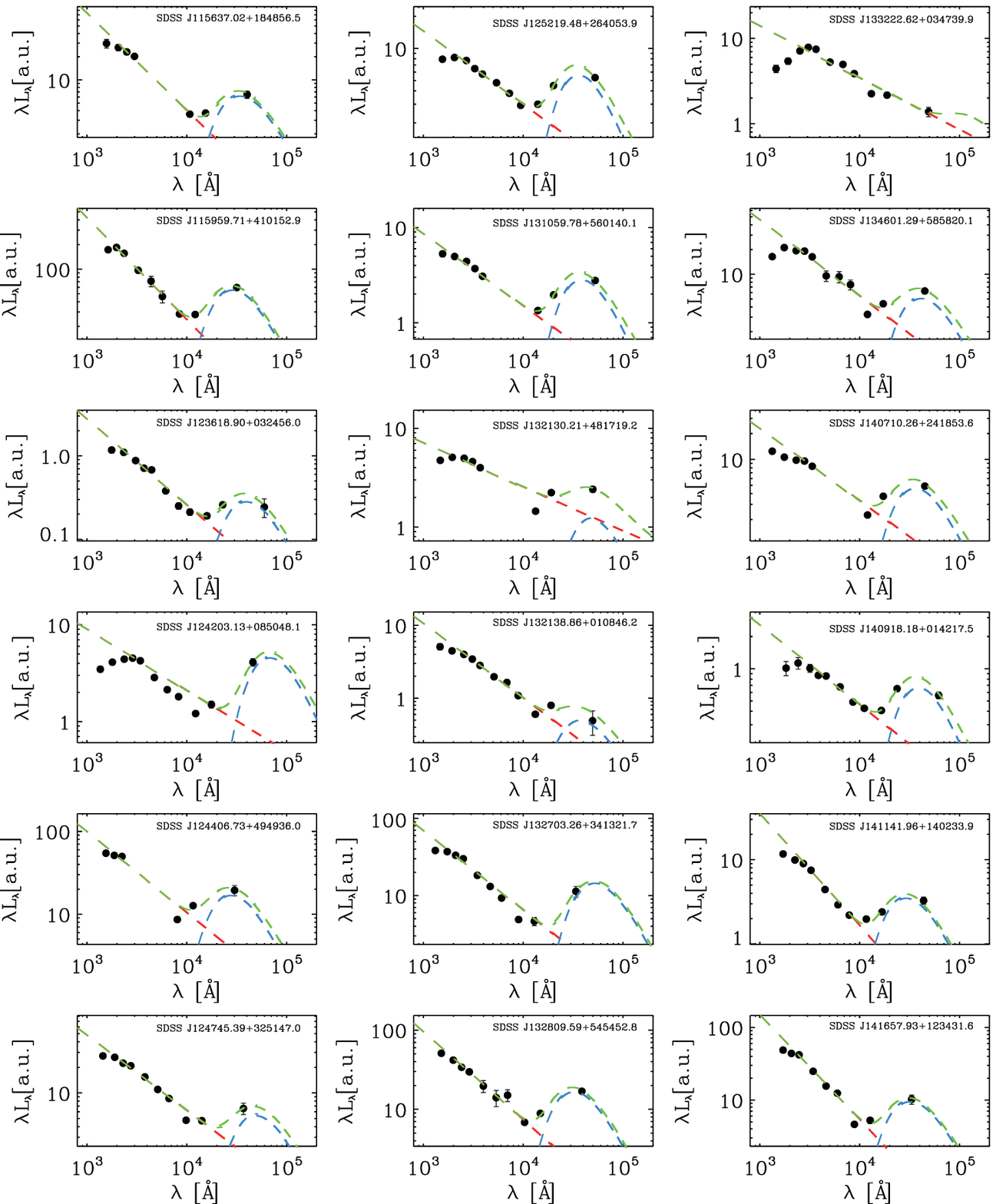}

{{\bf Figure~\ref{fig5}.} \textit {continued}}
\end{figure*}

\begin{figure*}[!htp]
\centering
\epsscale{.80}
\plotone{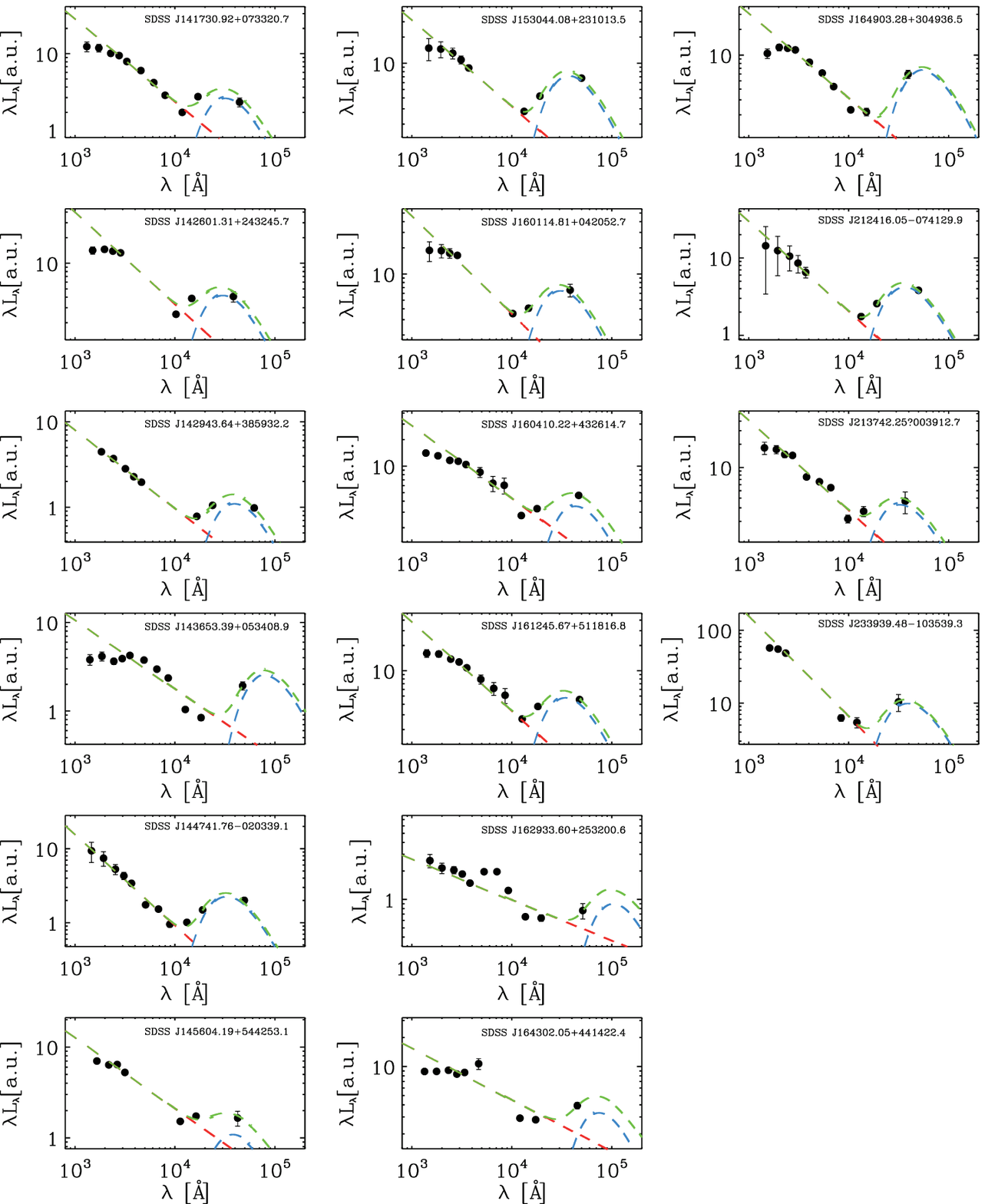}

{{\bf Figure~\ref{fig5}.} \textit {continued}}
\end{figure*}

\end{document}